# INCLUSION ANALYSIS

David Colver
Operis, 8 City Road, London EC1Y 2AA
dcolver@operis.com

**ABSTRACT**

*Inclusion analysis is the name given by Operis to a black box testing technique that it has found to make the checking of key financial ratios calculated by spreadsheet models quicker, easier and more likely to find omission errors than code inspection.*

## 1    CALCULATE MORE THAN ONCE

One approach among many [Pryor, 2004] [Panko, 2006] for improving the chance that a spreadsheet delivers the results that it is intended to deliver is to calculate quantities two or more different ways [Ettema et al, 2001]. If independent methods produce identical answers, there is a good chance, though no guarantee, that those answers are trustworthy. If the answers are not identical, the act of reconciling them and fathoming out what is causing the difference is often illuminating, not just about what mistake has been made, but about the underlying problem addressed by the spreadsheet and the assumptions being made in solving it.

Calculating a result by an alternative method is a standard weapon of a spreadsheet auditor. Most of the recent development focus of the Operis Analysis Kit [Oak, 2010], a software application for developing and reviewing spreadsheets public by the author's own company Operis, has been directed at accelerating the production of parallel reconstructions and of identifying where and why they have diverged from the original model. At least one of the large international accountancy practices declines to perform spreadsheet audits [Croll, 2003], or does so only grudgingly, preferring to reconstruct the spreadsheet and reconcile the outputs. This process no doubt delivers outstanding results, but not every client values them enough to pay five times as much for them.

Spreadsheet developers don't have to wait for an outside auditor to subject their spreadsheet to parallel reconstruction. They can calculate results repeatedly themselves, in their own spreadsheet. An obvious example concerns a table of numbers.

TABLE 1

| | |
|---|---|
| T | Table of numbers |
| C | Column totals |
| R | Row totals |
| G | Grand total |
| A | All of the above. |

The table has rows totals, column totals, and a grand total. It therefore offers opportunities to calculate the table total four times over. There are some simple things that ought to be the case. To pick just two:

$$SUM(R) = SUM(C)$$

(author's favourite) $\quad SUM(A) = G * 4.$

Since these tests are trivial to code, a conscientious spreadsheet developer will include them as a matter of course. A well constructed spreadsheet can easily devote 25% of its formulae to checking that relationships that should hold do in fact do so.

## 2 FINANCIAL STATEMENTS ADD UP

Many spreadsheet models concern themselves with the making of projections of the future financial statements of ventures that participants are contemplating launching or funding. Those financial statements have a happy property: they add up. At least, they ought to add up. The first steps in reviewing an unfamiliar financial model are

- to locate the financial statements (often quite a bit harder to do than one might imagine)
- to fathom out which entity the statements purport to relate to (again, not always crystal clear)
- because we are about to rely heavily on this property, to check that the financial statements add up.

Oversimplifying dramatically for clarity, we can say that both a cash flow forecast and a profit and loss account amount to a list of revenues and costs that result in a bottom line. The bottom line for the cash flow is some measure of cash generated, and that for the profit and loss statement is some measure of earnings retained. Symbolically

$$R - C = bl.$$

where R represents Revenue, C represents Costs, and bl is the relevant bottom line.

If we follow the convention that costs will always appear on the financial statements as negative numbers, then we can write

$$R + (C) = bl,$$

the brackets indicating that the costs are expressed as negatives. If now we break the list of revenues R into components $r_1$ to $r_4$, and the costs C into $c_1$ to $c_4$, we can write

$$r_1 + r_2 \ r_3 + r_4 + (c_1) + (c_2) + (c_3) + (c_4) + (bl) = 0.$$

Now we are free to regroup these terms in any clusters we like, for example

$$[\ r_1 + r_2 + (c_1) + (c_3)\ ] + [r_3 + r_4 + (c_2) + (c_4) + (bl)\ ] = 0.$$

Much the same arguments apply to a balance sheet. Because it balances,

$$A = L,$$

where A represents the assets and L represents the liabilities. And because it adds up too,

$$a_1 + a_2 + a_3 + a_4 + (l_1) + (l_2) + (l_3) + (l_4) = 0$$

where $a_1$ to $a_4$ are components of A, and $(l_1)$ to $(l_4)$ are components of L, expressed as negative numbers.

These items can be regrouped in arbitrary clusters, as before, such as

$$[\ a_1 + a_2 + (l_4)\ ] + [\ a_3 + a_4 + (l_1) + (l_2) + (l_3)\ ] = 0$$

## 3    INCLUSION ANALYSIS: SIMPLE EXAMPLE

These results are hardly profound enough to be placed on the current frontier of finance theory.  But they are useful in developing a simple method for testing the ratios that are often what can be considered the outputs that financial models finally deliver after many megabytes of striving.

Consider a simple project in which shareholders invest in the capital costs of a factory.  Some of the money it generates once up and running is delivered to the shareholders as dividends.  The rest is retained to cover closing costs, and to repay to the shareholders their initial investment.

TABLE 2

| Year | 2008 | 2009 | 2010 | 2011 | 2012 |
|---|---|---|---|---|---|
| Revenue |  | 80 | 80 | 80 |  |
| Costs |  |  |  |  |  |
|    construction | (60) |  |  |  |  |
|    operating |  | (20) | (20) | (20) |  |
|    decommissioning |  |  |  |  | (30) |
| Shareholders |  |  |  |  |  |
|    initial investment | 60 |  |  |  |  |
|    dividends |  | (30) | (30) | (30) |  |
|    return of capital |  |  |  |  | (60) |
| Increase in cash at bank | - | 30 | 30 | 30 | (90) |
| IRR |  |  |  |  |  |
|    to project | 83.93% |  |  |  |  |
|    to shareholders | 23.38% |  |  |  |  |

The ratios at the foot of the table are the outputs of the model to which most attention will be paid, and it is important to check carefully that they have been correctly calculated.  We will start with the project IRR.

The first step is to get the cash flow in such a form that it adds up to zero.  The costs are already expressed following the convention that, since they are outflows, they appear as negative numbers.  If this was not the case, we would invert them.  We do, though, need to invert the line "Increase in cash at bank", which is the equivalent in this example of what we referred to as bl, or bottom line, above.

After inverting the bottom line and inserting a row to check that the cash flow really does total zero, we get the following. (Shaded lines show what has changed.)

TABLE 3

| Year | 2008 | 2009 | 2010 | 2010 | 2012 |
|---|---|---|---|---|---|
| Zero check | 0 | 0 | 0 | 0 | 0 |
| Revenue |  |  | 80 | 80 | 80 |
| Costs |  |  |  |  |  |
|     Construction |  | (60) |  |  |  |
|     Operating |  |  | (20) | (20) | (20) |
|     Decommissioning |  |  |  |  | (30) |
| Shareholders |  |  |  |  |  |
|     initial investment |  | 60 |  |  |  |
|     dividends |  |  | (30) | (30) | (30) |
|     return of capital |  |  |  |  | (60) |
| Increase in cash at bank |  | - | (30) | (30) | (30) | 90 |
| IRR |  |  |  |  |  |
|     to project |  | 83.93% |  |  |  |
|     to shareholders |  | 23.38% |  |  |  |

If at this point we found that the "zero check" line did not show zeroes, we would know, either that some of the rows don't follow our chosen sign convention, or that the cash flow didn't add up. The latter possibility is more common than one might imagine. Sometimes large teams of highly paid bankers have worked on multi-billion dollar projects for a year or two without recognising this basic flaw in the projections upon which they have been relying.

Now we seek to reproduce the 83.93% IRR. That cell uses Excel's IRR function to calculate the rate of return implicit in a row of net cash flows which is not shown in this table, and is to be found elsewhere in the model. We follow the IRR formula back to its precedents and reproduce those at the top of our spreadsheet. We then recalculate the IRR from these cash flows and show that we can reproduce the reported return

TABLE 4

| Year | 2008 | 2009 | 2010 | 2011 | 2012 |
|---|---|---|---|---|---|
| Relevant cash flow | (60) | 60 | 60 | 60 | 0 |
| IRR |  |  |  |  |  |
|     calculated from above | 83.93% |  |  |  |  |
|     reported below | 83.93% |  |  |  |  |
|     discrepancy | 0.00% |  |  |  |  |
| Zero check |  | 0 | 0 | 0 | 0 | 0 |
| Revenue |  |  | 80 | 80 | 80 |  |
| Costs |  |  |  |  |  |
|     construction |  | (60) |  |  |  |
|     operating |  |  | (20) | (20) | (20) |
|     decommissioning |  |  |  |  | (30) |
| Shareholders |  |  |  |  |  |
| initial investment | 60 | | | | |
| dividends | | (30) | (30) | (30) | |
| return of capital | | | | | (60) |
| Increase in cash at bank | - | (30) | (30) | (30) | 90 |
| IRR | | | | | |
| to project | 83.93% | | | | |
| to shareholders | 23.38% | | | | |

Having established that we can reproduce the rate of return, we can be confident that the "relevant cash flow" is indeed the one that the model is using to calculate the project IRR. We now partition the elements of the original cash flow statement into two clusters, with the aim of showing which of the items form part of the "relevant cash flow, and which ones don't. This is quite easy to do by eye in this example, though that is not always the case.

TABLE 5

| Year | 2008 | 2009 | 2010 | 2011 | 2012 |
|---|---|---|---|---|---|
| Relevant cash flow | (60) | 60 | 60 | 60 | - |
| IRR | | | | | |
| calculated from above | 83.93% | | | | |
| reported below | 83.93% | | | | |
| discrepancy | 0.00% | | | | |
| Discrepancy | | | | | |
| included-relevant cash flow | 0 | 0 | 0 | 0 | 0 |
| included+excluded | 0 | 0 | 0 | 0 | 0 |
| Total of items | | | | | |
| Included | (60) | 60 | 60 | 60 | - |
| Excluded | 60 | (60) | (60) | (60) | - |
| ITEMS INCLUDED | | | | | |
| Revenue | | 80 | 80 | 80 | |
| Costs | | | | | |
| construction | (60) | | | | |
| operating | | (20) | (20) | (20) | |
| ITEMS EXCLUDED | | | | | |
| Costs | | | | | |
| decommissioning | | | | | (30) |
| Shareholders | | | | | |
| initial investment | 60 | | | | |
| dividends | | (30) | (30) | (30) | |
| return of capital | | | | | (60) |
| Increase in cash at bank | - | (30) | (30) | (30) | 90 |
| IRR | | | | | |
| to project | 83.93% | | | | |
| to shareholders | 23.38% | | | | |

The newly introduced lines, again shaded, serve merely to

- indicate which lines are in the included cluster and which in the excluded
- add up those included and excluded clusters
- verify that the included total matches the "relevant cash flow" at the top of the table
- verify that the included and excluded items combined add still add to zero, which is the duty that was previously performed by the "zero check" line.

Now we can interpret the result. The "Items included" section shows us that the project IRR quoted concerns a cash flow that is made up of some revenues less some costs. We could have learned that by looking at the formula used by the underlying model to calculate the "relevant cash flow"; but we didn't have to look at the formula to work out what must be in it.

A project IRR made up of revenues less various costs sounds plausible. But now we get a second bite of the cherry. We can look also at the items excluded, the ones that are on the cash flow statement but don't play any part in the project IRR. There are the various involvements of the shareholders; since they are the providers of finance, their participation is correctly excluded from a measure of the underlying project economics. But also excluded is the cost of decommissioning the plant at the end of its useful life. Leaving out the cost of abandoning the project is a material misstatement of the project's economics.

It would be hard, but possible, to examine the formula that derives the "relevant cash flow" and notice that some cost is missing, if that cost appears in nearly every financial projection. But not every financial model includes a decommissioning cost, and it would be very hard to look at the formula and notice that one relatively small and rather specialised cost is missing. Here, though, its exclusion is obvious.

The inclusion analysis demonstrates, in a prominent position near the top of the table, that the excluded items are equal in magnitude to the included ones. To this extent, it is exhaustive, in the sense that there can be no elements of the financial statement which are not considered for possible inclusion in the ratio.

## 4   ANOTHER EXAMPLE

Having tested the project IRR, we can repeat the process for the shareholders' IRR. All we have to do is

- copy the analysis already completed
- relink the line "relevant cash flow" so that it points to the source of the cash flows used in the equity return calculation
- verify that we can reproduce the shareholders' IRR from those cash flows
- move the lines around between the included and excluded sections until we can explain the new relevant cash flow.

TABLE 6

| Year | 2008 | 2009 | 2010 | 2011 | 2012 |
|---|---|---|---|---|---|
| Relevant cash flow | 60 | (30) | (30) | (30) | - |
| IRR | | | | | |
|   calculated from above | 23.38% | | | | |
|   reported below | 23.38% | | | | |
|   discrepancy | 0.00% | | | | |
| Discrepancy | | | | | |
|   included-relevant cash flow | 0 | 0 | 0 | 0 | 0 |
|   included+excluded | 0 | 0 | 0 | 0 | 0 |
| Total of items | | | | | |
|   included | 60 | (30) | (30) | (30) | - |
|   excluded | (60) | 30 | 30 | 30 | - |
| ITEMS INCLUDED | | | | | |
| Shareholders | | | | | |
|   initial investment | 60 | | | | |
|   dividends | | (30) | (30) | (30) | |
| ITEMS EXCLUDED | | | | | |
| Revenue | | 80 | 80 | 80 | |
| Costs | | | | | |
|   construction | (60) | | | | |
|   operating | | (20) | (20) | (20) | |
|   decommissioning | | | | | (30) |
| Shareholders | | | | | |
|   return of capital | | | | | (60) |
| Increase in cash at bank | - | (30) | (30) | (30) | 90 |
| IRR | | | | | |
|   to project | 83.93% | | | | |
|   to shareholders | 23.38% | | | | |

Here we can see that the cash flow whose rate of return is being tested is the dividends received by the shareholders, net of their initial investment. This looks reasonable enough. But it isn't until we inspect the items excluded from the ratio calculation that we notice that the eventual return of capital to the shareholders has been omitted from the ratio. It's one of the cash flows experienced by the shareholders and certainly belongs in the ratio.

## 5   BALANCE SHEET RATIOS

Exactly the same method can be used in tests involving balance sheet quantities. As we've seen, balance sheets conveniently sum to zero if the liabilities are expressed as negative numbers.

Consider, for example, the following simple balance sheet.

TABLE 7

| ASSETS | |
|---|---|
| Fixed assets | 100 |
| Current assets | |
|   cash | 1 |
| Total | 101 |
| | |
| LIABILITIES | |
| Debt | |
|   senior loan | 79 |
|   equity bridge loan | 10 |
|   shareholder loan | 5 |
| Equity | |
|   share capital | 5 |
|   retained earnings | 2 |
| Total | 101 |
| | |
| RATIOS | |
| Debt:equity ratio | 89% |

We wish to prove the debt:equity ratio, reported as 89%. As before, our first step is to restate the balance sheet so that it adds to zero. We invert the liabilities, remove the balance sheet footings and extraneous headings, and put in a test that it does adds to zero.

TABLE 8

| Zero check | 0 |
|---|---|
| Fixed assets | 100 |
| Current assets | |
|   cash | 1 |
| Debt | |
|   senior loan | (79) |
|   equity bridge loan | (10) |
| Equity | |
|   share capital | (10) |
|   retained earnings | (2) |
| RATIOS | |
| Debt:equity ratio | 89% |

Now we examine the debt:equity formula and seek to reproduce its result.

TABLE 9

| Ratio components | | |
|---|---|---|
|    top of fraction: debt | (A) | 79 |
|    bottom of fraction: debt + equity | (B) | 10 |
| Debt:equity ratio | | |
|    recalculated from above | (A)/(A+B) | 89% |
|    reported below | | 89% |
|    discrepancy | | 0% |
| Zero check | | 0 |
| Fixed assets | | 100 |
| Current assets | | |
|    cash | | 1 |
| Debt | | |
|    senior loan | | (79) |
|    equity bridge loan | | (10) |
| Equity | | |
|    share capital | | (10) |
|    retained earnings | | (2) |
| RATIOS | | |
| Debt:equity ratio | | 89% |

Now that we know what elements make up the ratio, we can seek to match those to the elements of the balance sheet, identifying what participates in the top of the fraction, what participates in the bottom, and what takes no part in it. The result is in Table 10, over the page.

As before, the items that are included in the ratio under discussion are interesting, but every bit as interesting are the items that play no part in the ratio. One of these is the equity bridge loan. Spotting that, we can ask immediately, if the senior loan considered to be debt for the purposes of this ratio, shouldn't the equity bridge loan be considered too? Or maybe it should be part of the equity, since it is standing in for investment that would otherwise be provided by the shareholders, and is being guaranteed by them?

There's a case to be made for either alternative, but not one for ignoring the bridge loan altogether. Leaving equity bridge loans out of ratios is the serious fault of the moment, which has almost derailed several recent deals. Like other errors of omission, it is spotted quickly and with certainty by an inclusion analysis, and is relatively hard to spot by simply studying the formulae.

While we are it at, we can ask whether the ratio is correct to consider the share capital as the only kind of equity, or whether it should not also address the retained earnings, which the inclusion analysis shows firmly to have been excluded from the calculation.

TABLE 10

| | |
|---|---:|
| Ratio components | |
|     top of fraction: debt | 79 |
|     bottom of fraction: debt + equity | 10 |
| Debt:equity ratio | |
|     recalculated from above | 89% |
|     reported below | 89% |
|     Discrepancy | 0% |
| Discrepancies | |
|     included in debt (below)+ top of fraction: debt (above) | 0 |
|     included in equity (below) + bottom of fraction: equity (above) | 0 |
|     included in debt + included in equity + excluded | 0 |
| Totals | |
|     included in debt | (79) |
|     included in equity | (10) |
|     excluded | 89 |
| **INCLUDED IN DEBT** | |
|     senior loan | (79) |
| **INCLUDED IN EQUITY** | |
|     share capital | (10) |
| **EXCLUDED** | |
| Fixed assets | 100 |
| Current assets | |
|     cash | 1 |
| Debt | |
|     equity bridge loan | (10) |
| Equity | |
|     retained earnings | (2) |
| RATIOS | |
| Debt:equity ratio | 89% |

This inclusion analysis is a little more sophisticated than the earlier ones because it considers the two elements of a ratio rather than a single quantity. The author terms them three-way inclusion analyses.

## 6   MISMATCHES

With luck, one can quickly find how the elements of a financial statement need to be partitioned between included and excluded amounts to reproduce the components of a ratio. Then all one has to do is to study the result, and work out whether the included items really belong inside the ratio, and (the key point of the analysis) the excluded items really belong outside.

But often there is no combination of the rows in the financial statement that will reproduce the sought-for numbers.

- One reason is that the model being tested is simply defective. It has marshalled the wrong items in assembling the ratio.

- Another reason is that the financial statements are too coarse-grained. For example, a model may correctly factor some costs into a ratio but not others. Under these conditions, it becomes necessary to replace the single operating costs line by its components, and to allocate those between the included and excluded groupings. All the time, the rule that the financial statement add up, and the fact that there is a test at the top of the analysis constantly monitoring what follows for this crucial property, are aids in ensuring that the decomposing of lines is done quickly and correctly.

- A further possible reason is that the model has correctly deviated from the financial statements, to reflect some fine detail of the transaction. Under these conditions it becomes necessary to add extra lines to the analysis to show what adjustments are necessary to match the reported ratios. It is rare that more than three or four lines of adjustment are necessary.

The three-way variant also makes it obvious when items appear on both the top and the bottom of a ratio. Measures of debt cover are particularly prone to this kind of double counting. They compare the cash available to service a business's debt with the cash needed to service that debt. Here, "service the debt" means pay the interest and repay the loan principal. Unfortunately, these aren't the only payments demanded by a bank. Banks also ask for all manner of impertinent fees. Are the fees part of the debt service, or part of the operating costs that are deducted in working out the cash available to service the debt? In all too many models, they appear in both, a fault that can be detected by formula inspection only by an auditor who knows what to look for, and who remembers to do it, but which can be detected by an employee with a few days' training through inclusion analysis.

## 7   WIDER USES OF INCLUSION ANALYSIS

Any situation where some, but not all, of the items in a group contribute to a result is a candidate for inclusion analysis. As just shown, financial ratios typical involve comparing some combination of lines on a financial statement with some other combination. But there are other applications too.

- Under many jurisdictions around the world, some costs are eligible for deductions in calculating the profits on which corporate taxes are levied, and some are not. An analysis showing which costs have been included in a tax calculation, and so treated as eligible for deductions, and which have been excluded from participation in the calculation, and so treated as ineligible, can easily be presented as an inclusion analysis. It will also show which revenues are subject to tax and which are exempt from it.

- Similarly, Value Added Tax is applied in Europe to some revenues and costs but not others. An inclusion analysis will show which is which.

- Every input cell in a model can be partitioned between those that influence operating income or pre-tax cash flow, and those which don't. An inclusion analysis distinguishes the two and quickly shows if inputs have been provided for costs or revenues that don't make their through the model appropriately.

These kinds of inclusion analysis are less straight forward to perform than the ones illustrated in this paper, but they have the same power to detect errors of omission. The trick is to make a habit of doing them, so that they become routine with practice, part of the atmosphere and culture of the team. Operis has managed to institutionalise them. Every incoming analyst is taught inclusion analysis on his third day with the firm, and every model that is reviewed has inclusion analysis applied to its inputs, to every ratio it produces, and to every tax calculation during its first few hours of review.

## 8   RELEVANCE TO SPREADSHEET TESTING

Inclusion analysis is a useful technique. It is easy to do, and can be taught to analysts with relatively little experience. It is easy to check their work because it is easy to specify a standard layout for the analysis.

The process of partitioning the financial statement into included and excluded items can also be trivially automated, with about thirty lines of Visual Basic more than enough to rearrange, in a handful of seconds, the lines of a cash flow statement that sums to zero, so that it shows the split between included and excluded items, and presents the result in a prescribed form. (Or, in some cases, to show that no such rearrangement is possible, which is generally a sign of a fault in the model under test.) However, it is good for the soul and education of the consultant to do the work by hand, takes hardly any longer, and leaves him with a deeper connection to and familiarity with the model being investigated.

Inclusion analysis quickly exposes issues that are hard to find otherwise, particularly ones concerning mismatched timing (comparing 2008 equity to 2007 debt). The excluded part of the analysis is a list of all of the candidates for errors of omission, which are notoriously hard to detect.

In theory, the included part of the analysis gives no information that can't be gleaned by examining the spreadsheet formula. However, it is not uncommon for a cost to be deducted once in one formula, only to be deducted again in another formula further along the chain of calculation, or perhaps to be added back again, so that rather than being counted once, the cost is counted zero or two times. Such faults are very hard to detect by formula inspection but immediately obvious in the inclusion analysis. Frequently, when model authors are told that they have double counted or omitted a cost, they report that they can't see where the fault is. Nor can the producer of the inclusion analysis, without looking rather carefully, but he can know with certainty that it must be there somewhere.

Ventures often undergo changes as they proceed. A contract may be subject to an extension or variation, or a company may be refinanced or sold. Under such circumstances, it is common for a financial model to be updated, and for the changes to be subjected to audit. In this way, financial models are often audited formally several times during their lives, giving model auditors the opportunity to revisit their own work of years ago, or recheck work that other firms have performed. Any issues that surface ought to be confined to the portions of model that have changed, since previous audits should have unearthed all the problems that applied to prior versions of the spreadsheet. However, Operis has found that inclusion analysis almost always quickly unearths issues that had been missed by the earlier exercises.

All these are practical benefits. But inclusion analysis is thought provoking in an academic sense too. The best research indicates that the most reliable way to detect spreadsheet errors is to undertake formula inspection, ideally in pairs or groups. This conclusion is supported by careful testing and measurement, something valuable but rare in this field.

> *"Although many prescriptive techniques have been put forth, only code inspection has been tested experimentally and has proven to be both safe and effective".* [Panko, 2005]

An inclusion analysis delivers a test that is equivalent to the formula inspection. There is often a perfect mapping between the included elements and the formulae leading to the ratio, one that is made obvious if the formulae are expressed in terms of meaningful names rather than spreadsheet coordinates. At the same time inclusion analysis delivers a second test, the excluded element, that has no analogue in code inspection. We don't have tests and measurements to prove it, only the anecdotal experience just cited; but it seems unlikely that a pair of complementary tests, demonstrably exhaustive, can be outperformed by a single test equivalent to just one of the pair.

## 9    SUMMARY

Not every spreadsheet result involves summarising or comparing (in a ratio) quantities that are equivalent to some but not all of the lines in a financial statement or other table of numbers that adds up. But where a spreadsheet result does have these properties, the opportunity is available to check the calculation two different ways.

One of the ways will be equivalent to inspecting the formula. The alternative way amounts to examining the formula's mirror image, a statement of all the items that have not been chosen for inclusion in the ratio component. This amounts to a list of all the potential errors of omission from the calculation, which is valuable because omissions are by their nature hard to identify.